\documentclass[referee]{raa}            

\usepackage{graphicx,times}
\usepackage{natbib}
\usepackage{amssymb,amsmath}
\usepackage{mathrsfs}
\usepackage{epstopdf}
\usepackage{longtable}
\usepackage{booktabs}
\usepackage{overpic}
\usepackage{xcolor}
\usepackage{url}

\newbox\grsign \setbox\grsign=\hbox{$>$} \newdimen\grdimen \grdimen=\ht\grsign
\newbox\laxbox \newbox\gaxbox
\setbox\gaxbox=\hbox{\raise.5ex\hbox{$>$}\llap
     {\lower.5ex\hbox{$\sim$}}}\ht1=\grdimen\dp1=0pt
\setbox\laxbox=\hbox{\raise.5ex\hbox{$<$}\llap
     {\lower.5ex\hbox{$\sim$}}}\ht2=\grdimen\dp2=0pt

\graphicspath{{figs/}}

\definecolor{malachite}{rgb}{0.34, 0.7, 0.22}

\begin{document}

\title{Morphological Statistic of Milky Way-like Galaxies}

\volnopage{Vol.0 (20xx) No.0, 000--000}      
\setcounter{page}{1}                         

\author{Zehao lin\inst{1}
\and Junye Wei\inst{1,2}
\and Ye Xu\inst{1,2}
\and Yingjie li\inst{1}
\and Chaojie Hao\inst{1}
\and Dejian Liu\inst{3}
\and Yiwei Dong\inst{1,2}
\and zhen-yu Wu\inst{4,5}
}

\institute{Purple Mountain Observatory, Chinese Academy of Sciences, Nanjing 210008, China; {\it linzh@pmo.ac.cn}\\
\and School of Astronomy and Space Science, University of Science and Technology of China, Hefei 230026, PR China\\
\and Three Gorges University, Yichang 443002, China\\
\and National Astronomical Observatories, Chinese Academy of Sciences, 20ADatun Road, Chaoyang District, Beijing 100101, PR China\\
\and School of Astronomy and Space Science, University of Chinese Academy of Sciences, Beijing 101408, PR China\\
\vs\no
{\small Received 20xx month day; accepted 20xx month day}
}

\abstract{
We analyze the spiral morphology of 1,738 nearby galaxies with similar Hubble type of the Milky Way ($i<60^\circ$, $z<0.048$) using multi-band optical images from the Sloan Digital Sky Survey and Digitized Sky Survey. Multiple-arm systems dominate, comprising $\sim$50-60\% of the sample and $\sim$75\% of large galaxies ($d_{\rm max} \geqslant 18~\mathrm{kpc}$). In $90\pm3$\% of multiple-arm galaxies, two inner arms extend into multiple outer arms; three-arm inner structures occur in $9\pm2$\%, and four or more are rare ($<1$\%). Morphology and inner-arm counts are consistent across bands at $\sim$80\%. Bifurcation points, marking the transition from two inner arms to multiple outer arms, occur in up to $\sim$40\% of galaxies, are more prominent in bluer bands tracing younger stellar populations, and cluster within 4-7~kpc of galaxy centers. These statistics suggest that the Milky Way's inner two-arm and outer multiple-arm morphology, including bifurcations at similar radii, is broadly consistent with that of a typical multiple-arm spiral galaxy.
}

\authorrunning{Z. lin et al.}
\titlerunning{Morphological Statistic of MW-like Galaxies}
\maketitle

\section{Introduction}
\label{intro}

In the 1840s, William Parsons constructed the 72-inch ``Leviathan of Parsonstown'' telescope at Birr Castle in Ireland. This instrument revealed spiral arm patterns in many bright Herschel and Messier ``nebulae'', marking the beginning of intriguing discoveries in galaxy morphology. Morphology serves as the logical starting point for understanding galaxies, with galaxy morphology being intrinsically linked to their star formation histories~\citep{Buta2013}. For nearly a century, the study of galaxy morphology has been a vital tool in understanding of the universe, and even today, it remains a cornerstone of galaxy research.

\cite{Hubble1926,Hubble1936} originally proposed the iconic Hubble sequence based on the combination of central bulge prominence and spiral arm pitch angles, later expanded by \cite{Vaucouleurs1959,Sandage1961}. However, the Hubble sequence lacks detailed classification of spiral arm morphology in disk galaxies. Spiral arm structures exhibit complex morphologies characterized by multiple parameters: amplitude (arm-interarm contrast), width, pitch angle, and the number of spiral arms. \cite{Elmegreen_Elmegreen1982} analyzed the spiral arm morphology of 305 disk galaxies and classified them into 12 subclass. Subsequent studies by \cite{Elmegreen_Elmegreen1987} established the prevailing tripartite classification: flocculent galaxy (F), multiple-arm galaxy (M), and grand-design galaxy (G). Grand-design arms typically manifest as a pair of symmetric, continuous spiral arms consistent with density-wave theory predictions; flocculent arms consist of short, fragmented nebulae and star-forming clumps likely originating from local gravitational instabilities or transient spiral arms~\citep{Dobbs_Baba_2014}; while multiple-arm galaxy is intermediate between them, exhibiting asymmetric branched or overlapping patterns, and challenging traditional monolithic theoretical frameworks. This morphological diversity suggests that spiral arm formation and maintenance mechanisms may diverge significantly depending on galactic mass, gas fraction, and environmental conditions~\citep{Kennicutt1998}.

The spiral structure of the Milky Way (MW) has long been recognized as a key probe of galactic dynamics. The classical four-arm model of \citet{Georgelin_Georgelin1976}, derived from optical and radio observations of young stellar tracers, established the so-called ``standard'' picture despite large kinematic distance uncertainties. Advances in VLBI parallax measurements have since enabled precise mapping of high-mass star-forming regions, culminating in the detailed spiral map of \citet{Reid+2019}, which revealed four principal arms along with numerous spurs and segments. These results, together with evidence for the Local Arm and localized irregularities, indicate that the MW departs from a symmetric grand-design morphology. Integrating maser parallaxes, {\it Gaia} DR3 massive star astrometry, and arm tangents, \citet{Xu+2023} proposed a new Milky Way model that, rather than extending all arms continuously around the disk as in classical four-arm model, introduces a conceptually distinct morphology: a multiple-arm galaxy with a symmetric two-arm inner pattern bifurcating into several outer arms. This reinterpretation, which remains under debate, provides a direct motivation to examine whether such a configuration is common among external spirals.

Extragalactic statistical patterns provide critical perspectives for reevaluating MW model. Although many studies have statistically analyzed the spiral arm morphology of spiral galaxies~\citep[e.g.,][]{Elmegreen_Elmegreen1987, Ann_Lee2013, Elmegreen+2011, Buta+2015}, these statistics have not detailed features such as the number of inner arms or the presence of spiral arm bifurcation points, which are critical for distinguishing between different MW models. Recently, we conducted a statistical analysis of the morphology of as many as over 5,000 nearby blue spiral galaxies, where an important concern is the number of inner/outer spiral arm and the existence and position of spiral arm bifurcation points~\citep[][]{Wei+2024}. The results reveal that 82\% of spiral galaxies exhibit inner two-arm structure, while 74\% of barred spiral galaxies show a configuration with two inner arms and multiple outer arms. Notably, no barred spiral galaxies display a continuous four-arm structure extending from the inner to outer regions.

It remains under-studied of key attributes of galaxies with similar Hubble type of the MW~\citep[hearafter MW-like galaxies,][]{Blitz+1983, Gerhard_2002, binney+2008} such as the debated number of inner arms in MW model~\citep{Drimmel_Spergel_2001}, the existence of bifurcation points, and their galactocentric distances. This study utilizes multi-band optical data from the Sloan Digital Sky Survey~\citep[SDSS,][]{York2000} and Digitized Sky Survey~\citep[DSS,][]{Lasker+2008} to conduct a census of nearby MW-like galaxy images, derive statistical patterns of MW-like galaxies, and provide a unique perspective for reexamining the MW models. Section~\ref{sec2} describes the selection and classification methods for galaxies. Section~\ref{sec3} presents statistical results of galaxies. Section~\ref{sec4} compares these results with MW models. Conclusions are provided in Section~\ref{sec5}.

Throughout this paper, we assume a Hubble constant of 70 km s$^{-1}$ Mpc$^{-1}$.

\section{Sample and method}
\label{sec2}

\subsection{Sample}
\label{sec2_1}

Given the prevailing consensus that the MW is a barred spiral galaxy with a likely Hubble type between Sb and Sc~\citep{Blitz+1983, Gerhard_2002, binney+2008}, this study focuses on comparison galaxies with Hubble classifications of SBb, SABb, SBbc, SABbc, SBc, and SABc. To ensure morphological clarity and statistical reliability, we adopt the following selection criteria:

\begin{enumerate}
    \item \textbf{Inclination angle less than 60 degrees} (i.e., an axial ratio smaller than 2), to limit the sample to galaxies that are sufficiently face-on and thereby minimize projection effects.
    
    \item \textbf{Minor axis length of at least 10 arcseconds}, ensuring that the galaxy spans a sufficient number of pixels for a reliable morphological analysis.
    
    \item \textbf{Redshift $z < 0.048$}, corresponding to a physical scale of better than 1 kpc per arcsecond, to avoid including distant galaxies that appear overly compact or blurred due to resolution limitations.
\end{enumerate}

Finally, we selected a total of 1,738 galaxies from the SIMBAD database, with their properties listed in Appendix Table~\ref{tab:galaxies}. While the selection criteria described above inevitably involve some degree of subjectivity, we have explored the sensitivity of our statistical results to variations in these constraints. This allows us to estimate how our conclusions might change if the sample were extended beyond the current selection limits. Future availability of higher-quality imaging data is expected to enable more comprehensive and robust analyses.

The galaxy images used in this study are drawn from the optical imaging data of the Sloan Digital Sky Survey~\citep[SDSS;][]{York2000} in the $g$-, $r$-, and $i$-bands, and from the Digitized Sky Survey~\citep[DSS;][]{Lasker+2008} in the blue and red bands. The high-resolution SDSS images enable pixel-level analyses of spiral arm structures, while the broader sky coverage of DSS effectively mitigates potential selection biases that could arise from the absence of SDSS imaging for certain galaxies.

\subsection{Image Preprocessing}
\label{sec2_2}

Before performing the statistical classification of spiral structures, we corrected the images for projection effects using a simple geometric deprojection. Since our sample galaxies are already largely face-on, this correction only introduces minor visual changes, but it provides a consistent basis for measuring the distances from bifurcation points to galaxy centers.

The ellipse fitting was performed only to determine the position angle of each galaxy on the sky, rather than to measure its axis ratio. Specifically, for each DSS or SDSS image, we extracted a square data centered on the galaxy, with a half-width equal to 1.5 times the semi-major axis listed in SIMBAD. For a nearly face-on galaxy, the projected galaxy region can be roughly approximated as a circle with radius equal to the semi-major axis, occupying a fractional area of ($\pi/3^2 \simeq 35\%$) within the data. Therefore, as a simple empirical criterion, we retained the brightest ($\sim35\%$) of pixels, corresponding to masking pixels with intensities below the 65th percentile of the intensity distribution in the data. The remaining pixels were then converted into a binary image for subsequent ellipse fitting. We then selected the largest connected structure in the binary image, since foreground stars and background sources generally produce much smaller structures than the target galaxy. Ellipses were fitted to this main structure, and the resulting position angle was adopted as the orientation of the galaxy.

\begin{figure}[!ht]
	\centering
	\includegraphics[height=0.15\textheight]{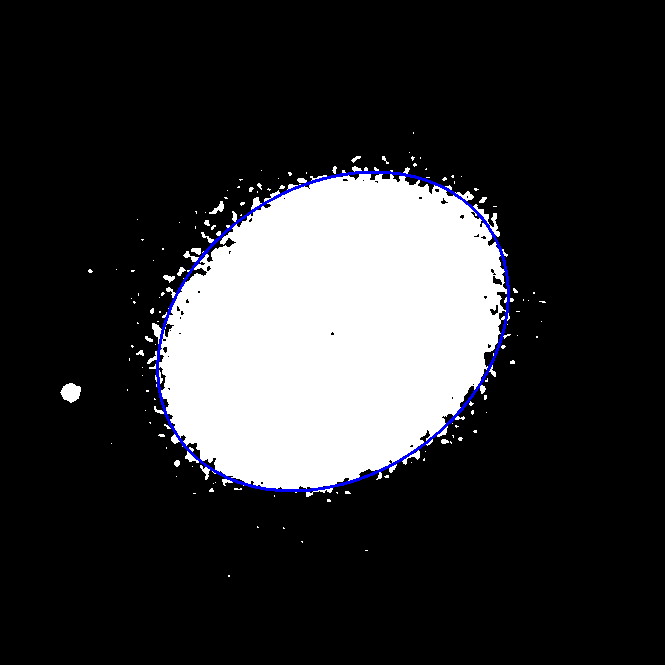}
	\includegraphics[height=0.15\textheight]{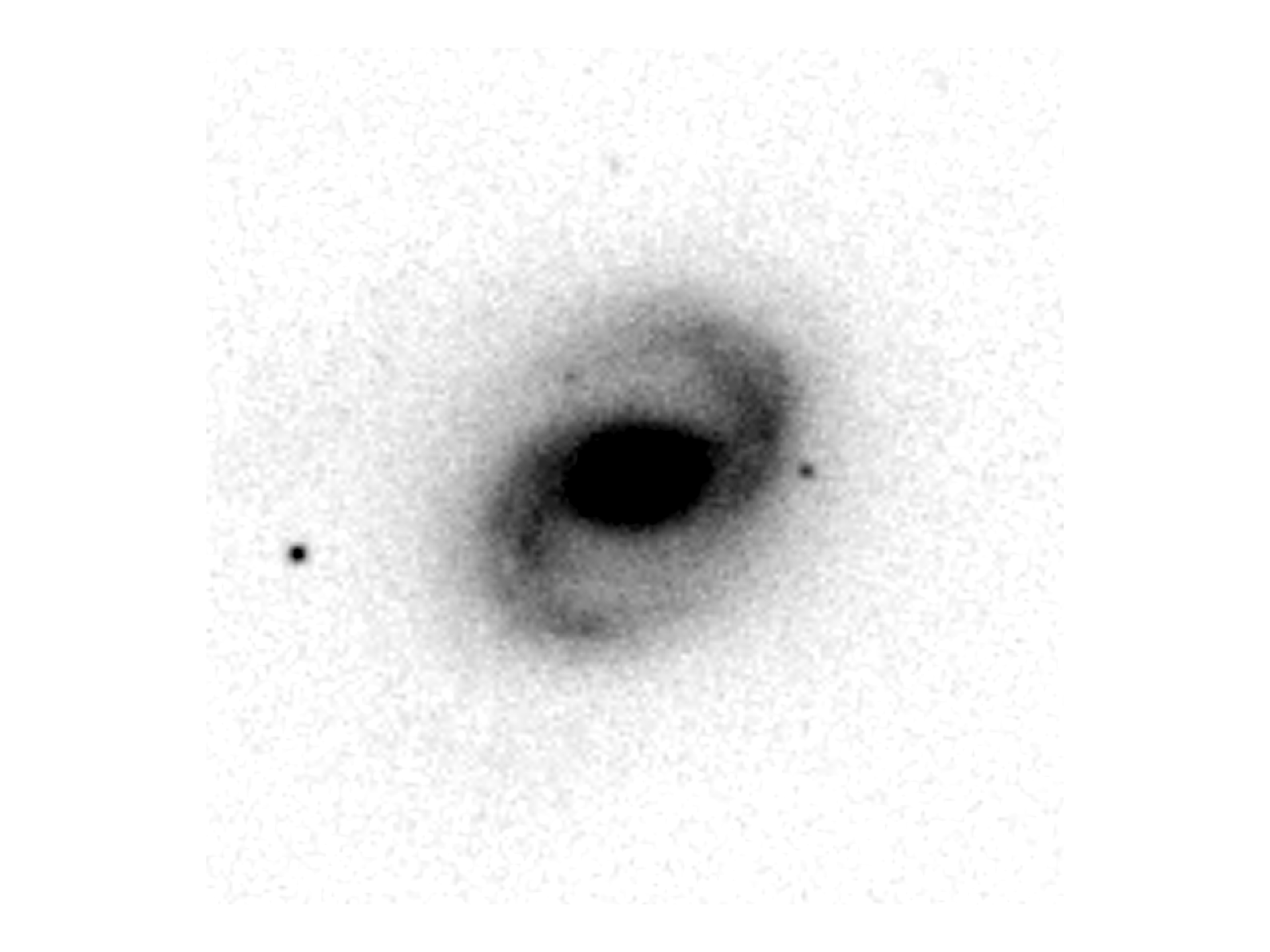}
	\includegraphics[height=0.15\textheight]{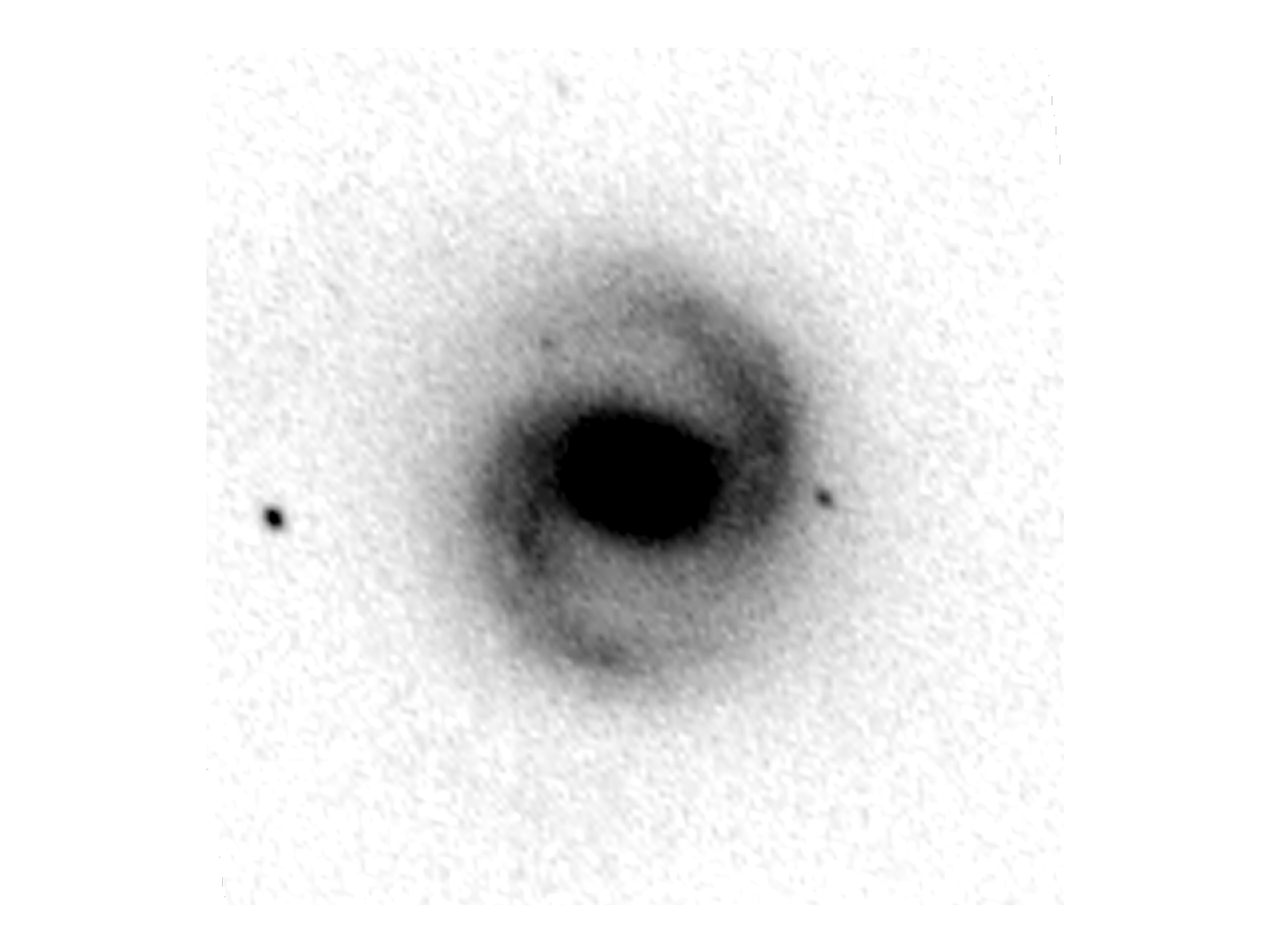}
	\caption{NGC 5431: Elliptical fitting (left panel), original image (middle panel), and deprojected image (right panel).}
	\label{fig:face_on}
\end{figure}

For the deprojection, the semi-major and semi-minor axes were taken from SIMBAD. After rotating the image according to the fitted position angle, we stretched the minor-axis direction by the ratio of the semi-major to semi-minor axis to approximate a face-on view. Figure~\ref{fig:face_on} illustrates this procedure: the left panel shows the elliptical fit to NGC~5431, the middle panel displays the original image, and the right panel presents the deprojected image adjusted to a nearly face-on orientation. We note that no luminosity-weighted radial profile fitting or radius-dependent isophotal modeling was performed, because the deprojection is used only as a first-order geometric correction for our nearly face-on sample.

\begin{figure}[!ht]
	\centering
	\includegraphics[width=0.75\textwidth]{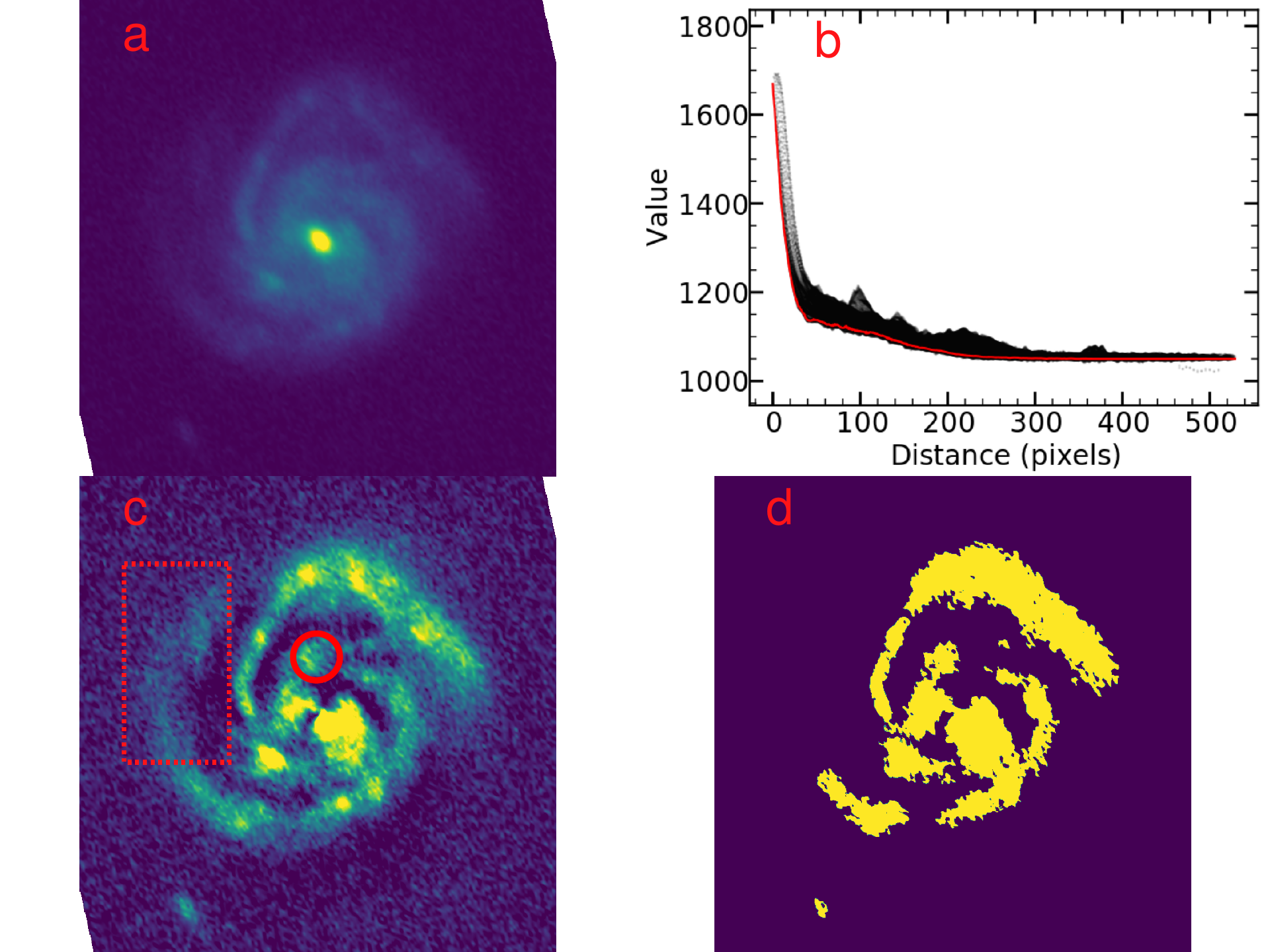}
	\caption{The reference images for spiral arm morphology classification, showing the MCG+08-24-035 in the $g$-band. The red dashed box highlights a region where a faint spiral arm structure becomes more prominent after background subtraction. The red circle marks the potential bifurcation point within the galaxy.}
	\label{fig:check_sample}
\end{figure}

Relative to the asymmetric spiral arms and bars, the nearly circularly symmetric disks and bulges dominates the contribution to the surface brightness of galaxies. Galactic centers are typically much brighter than the outskirts, causing the outer structures to appear significantly fainter in the images (as shown in Figure~\ref{fig:check_sample}a). To minimize the impact of the overly bright bulge regions and the stellar disks, whose surface brightness generally decreases exponentially with radius, we fitted the one-dimensional radial surface brightness profile for each galaxy. Given that galactic stellar disks can exhibit various surface brightness profiles~\citep[see][and references therein]{Kruit_Freeman2011}, we determined, for each radial bin and each galaxy, the 5th percentile of surface brightness values, which we adopted as the local background level for subsequent analyses (as shown as the red line in Figure~\ref{fig:check_sample}b).

By computing the ratio of the surface brightness of galaxies to this background level, we normalized the pixel values across the entire image to a comparable scale (as shown in Figure~\ref{fig:check_sample}c). Following this transformation, the brightness is no longer dominated by the galactic center, allowing the spiral arm structures to be more prominently revealed. This effect is particularly evident in the leftmost spiral arm segment highlighted by the red box in Figure~\ref{fig:check_sample}c, which was barely distinguishable in the original face-on image (Figure~\ref{fig:check_sample}a). Subtracting the background significantly enhances the visibility of asymmetric features such as spiral arms, thereby improving the robustness of our statistical analyses.

To facilitate the inspection process, we have highlighted regions in Figure~\ref{fig:check_sample}c where the surface brightness exceeds three times the standard deviation of the entire image (see Figure~\ref{fig:check_sample}d). It should be noted that this thresholding highlights only the strongest features, causing fainter structures to be omitted. For example, the leftmost spiral arm segment visible in Figure~\ref{fig:check_sample}c does not appear in Figure~\ref{fig:check_sample}d. Nevertheless, Figure~\ref{fig:check_sample}d effectively delineates the spiral arms and interarm regions, thereby enhancing the visual identification of these features. Accordingly, we use this image as an auxiliary tool during our statistical analysis and incorporate it into our inspection workflow.

\begin{figure}[!ht]
	\centering
	\includegraphics[width=0.8\textwidth]{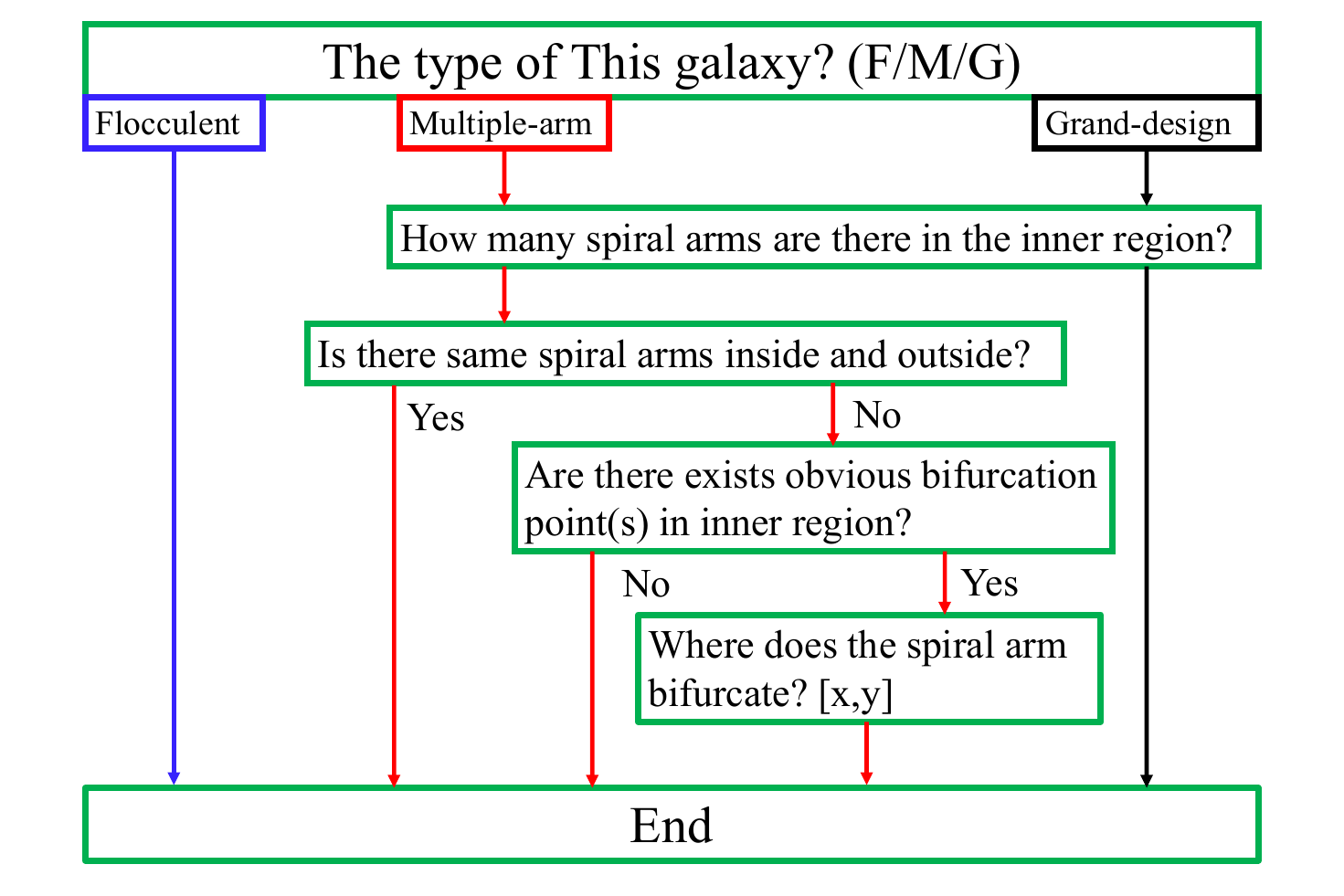}
	\caption{Flowchart for classifying spiral arm morphology.}
	\label{fig:check_flowchat}
\end{figure}

\subsection{Statistical procedure}
\label{sec2_3}

We analyzed each galaxy image following the workflow illustrated in Figure~\ref{fig:check_flowchat}, incorporating information from three different image versions (e.g., Figures~\ref{fig:check_sample}a, \ref{fig:check_sample}c, and \ref{fig:check_sample}d). 
As an example, the galaxy MCG+08-24-035 in the $g$-band (Figure~\ref{fig:check_sample}) displays two prominent spiral arms, one of which branches at the location indicated by the red circle in Figure~\ref{fig:check_sample}c. Based on this structure, we classify MCG+08-24-035 in the $g$-band as a multiple-arm galaxy with two inner arms. By identifying bifurcation regions and measuring their radial separation from the image center, we derived the galactocentric distances of spiral arm bifurcation points. 

Following the same procedure, we performed a statistical analysis on 1,738 galaxies, encompassing a total of 7,502 images across multiple bands. The classification results are summarized in Table~\ref{tab:galaxies}. Each image was independently inspected by two authors to ensure the objectivity and consistency of the classifications. The independent statistical differences between the two authors are detailed in Appendix~\ref{secb}. To further reduce potential bias, we randomized the order of the images during the analysis, ensuring that images of the same galaxy in different bands were not evaluated consecutively.

\section{Results}
\label{sec3}

\subsection{Morphological statistics}\label{sec3_1}

For all 1,738 galaxies in our sample, DSS blue- and red-band images are available and were used for morphological inspection. Among them, 1,342 galaxies also have high-resolution SDSS images in the $g$-, $r$-, and $i$-bands, which were used to provide more detailed structural information. For the remaining 396 galaxies without SDSS coverage, the detailed structural classifications were based on the DSS blue- and red-band images. Based on these imaging data, the statistical results of the spiral-arm morphology classifications are summarized in Table~\ref{tab:statis_result}.

\begin{table*}[!ht]
	\small
	\caption{Statistical results on spiral arm morphologies in different bands, with (A) and (B) representing different authors}
	\label{tab:statis_result}
	\hspace{+2.5cm}
	\begin{tabular}{|l|ccc|ccc|c|} 
		\hline
		image & F (A) & M (A) & G (A) &  F (B) & M (B) & G (B) & Total \\
		\hline
		DSS blue &  753 &  930 &  55 &  755 &  926 &  57 & 1738 \\ 
		DSS red  &  808 &  875 &  55 &  837 &  855 &  46 & 1738 \\ 
		SDSS g   &  471 &  810 &  61 &  452 &  852 &  38 & 1342 \\
		SDSS r   &  468 &  803 &  71 &  442 &  855 &  45 & 1342 \\
		SDSS i   &  518 &  752 &  72 &  505 &  801 &  36 & 1342 \\
		\hline
	\end{tabular}
\end{table*}

For the full sample of 1,738 galaxies inspected using DSS images, the two independent classifications give consistent spiral-arm morphology fractions. In the DSS blue band, flocculent galaxies account for 43 $\pm$ 1\%, multiple-arm galaxies for 54 $\pm$ 1\%, and grand-design galaxies for 3 $\pm$ 1\%. In the DSS red band, the corresponding fractions are 47 $\pm$ 1\%, 50 $\pm$ 1\%, and 3 $\pm$ 1\%, respectively. Here the quoted uncertainties represent the half-differences between the two independent classifications.

For the 1,342 galaxies analyzed with SDSS data: in the $g$-band, flocculent galaxies comprise $34 \pm 1$\%, multiple-arm galaxies $62 \pm 2$\%, and grand-design galaxies $4 \pm 1$\%. In the $r$-band, flocculent galaxies represent $34 \pm 1$\%, multiple-arm galaxies $62 \pm 2$\%, and grand-design galaxies $4 \pm 1$\%. In the $i$-band, flocculent galaxies account for $38 \pm 1$\%, multiple-arm galaxies $58 \pm 2$\%, and grand-design galaxies $4 \pm 1$\%. The results obtained from SDSS-based classifications are consistent with those of previous studies on spiral galaxy morphologies~\citep{Wei+2024}.

Table~\ref{tab:diference_vs_band} summarizes the discrepancies in spiral arm morphology classifications across different photometric bands. Overall, the majority of galaxies show consistent classifications between the DSS blue and red bands, with a consistency fraction of 87 $\pm$ 1\% based on the two independent classifications. Among the galaxies with band-dependent classifications, 9 $\pm$ 1\% of the full sample show a shift toward more flocculent morphologies in the red band (e.g., M~$\rightarrow$~F, G~$\rightarrow$~M, or G~$\rightarrow$~F). In contrast, 4 $\pm$ 1\% show a shift toward more regular spiral-arm morphologies (e.g., F~$\rightarrow$~M, M~$\rightarrow$~G, or F~$\rightarrow$~G).

\begin{table*}[!ht]
	\small
	\centering
	\caption{Statistical differences in spiral arm morphologies of galaxies in different bands}
	\label{tab:diference_vs_band}
	\begin{tabular}{|c|c|c|c|c|c|c|c|c|} 
		\hline
		   & consistent & inconsistent &$\rm{F}\rightarrow\rm{M}$ & $\rm{M}\rightarrow\rm{F}$ & $\rm{M}\rightarrow\rm{G}$ & $\rm{G}\rightarrow\rm{M}$ & $\rm{F}\rightarrow\rm{G}$ & $\rm{G}\rightarrow\rm{F}$  \\
		\hline
		DSS blue $\rightarrow$ DSS red (A) & 1495 & 243 &  84 & 137 &  11 &   9 &   0 &   2\\
		DSS blue $\rightarrow$ DSS red (B) & 1519 & 219 &  60 & 136 &   5 &  10 &   1 &   7\\
		\hline
		SDSS $g$ $\rightarrow$ SDSS $r$ (A)    &1028 & 314 & 119 & 113 &  45 &  32 &   1 &   4\\
		SDSS $g$ $\rightarrow$ SDSS $r$ (B)    &1201 & 141 &  63 &  55 &  13 &   8 &   2 &   0\\
		SDSS $r$ $\rightarrow$ SDSS $i$ (A)    &1014 & 328 & 102 & 151 &  35 &  33 &   3 &   4\\
		SDSS $r$ $\rightarrow$ SDSS $i$ (B)    &1173 & 169 &  44 & 104 &   6 &  12 &   0 &   3\\
		SDSS $g$ $\rightarrow$ SDSS $i$ (A)    &1009 & 333 & 107 & 149 &  44 &  28 &   0 &   5\\
		SDSS $g$ $\rightarrow$ SDSS $i$ (B)    &1169 & 173 &  51 & 102 &   9 &   9 &   0 &   2\\
		\hline
		DSS blue $\rightarrow$ SDSS $g$ (A)    & 989 & 353 & 192 & 106 &  28 &  16 &  10 &   1\\
		DSS blue $\rightarrow$ SDSS $g$ (B)    &1074 & 268 & 194 &  45 &   7 &  14 &   7 &   1\\
		DSS blue $\rightarrow$ SDSS $r$ (A)    &1005 & 337 & 184 &  94 &  37 &  14 &   8 &   0\\
		DSS blue $\rightarrow$ SDSS $r$ (B)    &1069 & 273 & 199 &  40 &  13 &  13 &   7 &   1\\
		DSS blue $\rightarrow$ SDSS $i$ (A)    &1001 & 341 & 163 & 122 &  36 &  11 &   8 &   1\\
		DSS blue $\rightarrow$ SDSS $i$ (B)    &1062 & 280 & 171 &  74 &  11 &  19 &   5 &   0\\
		DSS red  $\rightarrow$ SDSS $g$ (A)    & 975 & 367 & 212 &  98 &  28 &  18 &  11 &   0\\
		DSS red  $\rightarrow$ SDSS $g$ (B)    &1055 & 287 & 226 &  36 &   9 &   8 &   8 &   0\\
		DSS red  $\rightarrow$ SDSS $r$ (A)    &1006 & 336 & 199 &  82 &  32 &  12 &  11 &   0\\
		DSS red  $\rightarrow$ SDSS $r$ (B)    &1044 & 298 & 231 &  33 &  15 &   9 &  10 &   0\\
		DSS red  $\rightarrow$ SDSS $i$ (A)    &1002 & 340 & 176 & 108 &  34 &  12 &  10 &   0\\
		DSS red  $\rightarrow$ SDSS $i$ (B)    &1054 & 288 & 202 &  63 &   9 &   8 &   6 &   0\\
		\hline
	\end{tabular}
\end{table*}

For galaxies analyzed using SDSS $g$-, $r$-, and $i$-band images, 81~$\pm$~6\% maintain consistent morphologies across bands. A fraction of galaxies (11~$\pm$~3\%) exhibit a trend toward more flocculent morphology when progressing from bluer to redder bands ($g$$\rightarrow$$r$, $r$$\rightarrow$$i$, or $g$$\rightarrow$$i$), while 8~$\pm$~3\% show the opposite trend toward more regular structures .

These findings suggest that spiral arm morphologies are generally robust across different bands for most galaxies. However, approximately $\sim$20\% of galaxies exhibit band-dependent variations. The magnitude of these variations is comparable to those observed in inter-author classification differences or reclassifications over time~\citep{Wei+2024}. Notably, there is a slight tendency for galaxies to appear more flocculent in redder bands, indicating a possible wavelength-dependent visibility of spiral structure. This apparent inconsistency with the results of \citet{Buta+2015} and \citet{Elmegreen+2011} may be explained by weaker spiral arm features in red bands having lower contrast relative to their surroundings, making them more easily overlooked.

Spiral arms generally appear sharper and more distinct in bluer bands, while exhibiting a more diffuse and extended morphology in redder bands. This is consistent with the fact that spiral arms trace regions of enhanced star formation, which are more prominently detected in ultraviolet and blue optical bands~\citep{Masters2010, Mosenkov+2024}. \citet{Yu+2018} similarly reported systematically higher spiral strength measurements in bluer bands, and \citet{Savchenko+2020} observationally confirmed that the arm-to-total light ratio peaks in the $g$-band. Consequently, the enhanced visibility of spiral structures in bluer bands contributes to the observed statistical trend of galaxies appearing more flocculent in morphology when transitioning from blue to red bands. 

For the 1,342 galaxies with SDSS coverage, we further compared the classifications based on DSS images with those based on SDSS $g$-, $r$-, and $i$-band images. Averaging over all DSS--SDSS band combinations and the two independent classifications, 77 $\pm$ 3\% of the galaxies show consistent classifications. Among the remaining galaxies, 6 $\pm$ 3\% show a shift toward more flocculent morphologies in the SDSS images, whereas 17 $\pm$ 2\% show a shift toward more regular spiral-arm morphologies. 
This asymmetric trend is seen across the SDSS $g$, $r$, and $i$-bands, indicating that it is mainly caused by the higher spatial resolution and sharper PSF of the SDSS images, which can reveal more continuous or coherent spiral-arm structures, rather than by a random band-dependent classification bias.

\subsection{Influence of Physical and Observational Parameters on Morphology statistics}\label{sec3_2}

For a more detailed investigation, we select galaxies that exhibit consistent spiral arm classifications across different bands as the morphologically consistent subsample (hereafter MC subsample). Considering the higher spatial resolution and sharper PSF of the SDSS images, we adopted the SDSS-based classifications whenever SDSS data are available; DSS-based classifications were retained only for galaxies lacking SDSS coverage.
This subsample includes a total of 1,043 galaxies—267 from DSS and 776 from SDSS—representing approximately 60\% of the full sample. Within this subsample, flocculent galaxies account for 383 (125 from DSS and 258 from SDSS, 37\%), multiple-arm galaxies comprise 645 (134 from DSS and 511 from SDSS, 62\%), and grand-design galaxies represent only 15 (8 from DSS and 7 from SDSS, 1\%).
The notably lower fraction of grand-design galaxies in MC subsample likely stems from the stricter requirements necessary for consistent classification across bands. Nonetheless, the overall distribution of morphological types remains statistically consistent with that of the full sample.

\begin{figure}[!ht]
	\centering
	\includegraphics[width=0.5\textwidth]{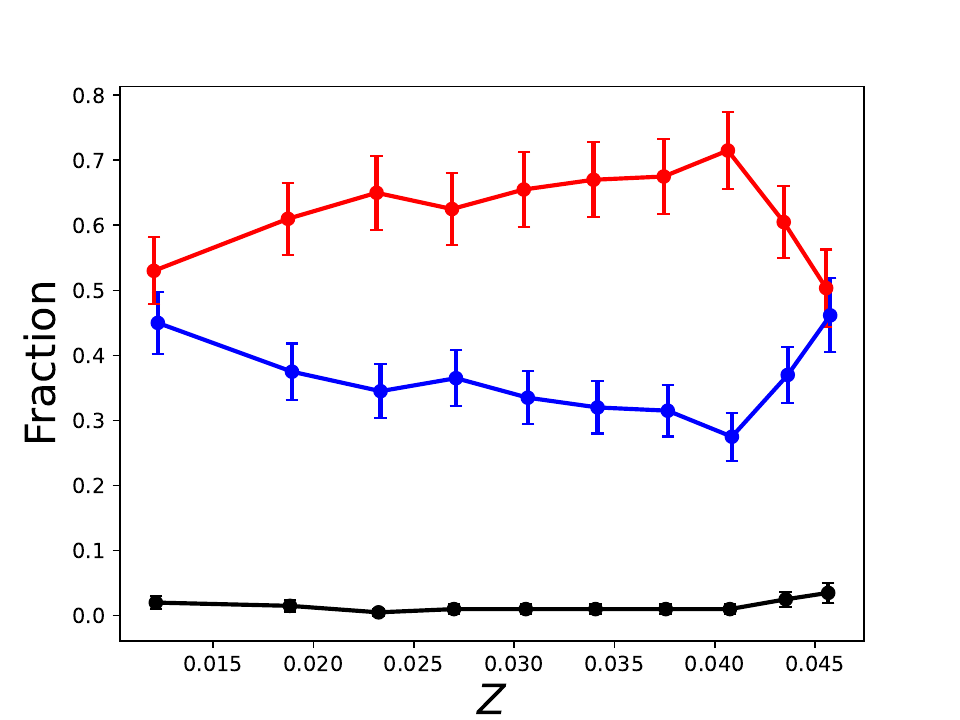}
	\caption{The galaxy morphologies distribution with redshift. Shown are flocculent galaxies (blue), multiple-arm galaxies (red), and grand-design galaxies (black).}
	\label{fig:redshift_morpholoty}
\end{figure}

The statistical results of spiral arm morphologies as a function of redshift are presented in Figure~\ref{fig:redshift_morpholoty}. In this analysis, galaxies are sorted by increasing redshift, and morphological statistics are computed for every set of 200 galaxies with a step size of 100. The error bars represent Poisson uncertainties, calculated as $\sqrt{n}$, where $n$ is the number of galaxies in each bin.
The results indicate that the distribution of spiral arm morphologies does not show significant variation with redshift. This finding is consistent with the results of \citet{Wei+2024}, who reported similar morphological distributions for spiral galaxies in the redshift range $0.03 < z < 0.085$. Building on their work, our analysis extends this conclusion down to lower redshifts, demonstrating that the statistical distribution of spiral arm morphologies remains stable from $z \sim 0$ to $z \sim 0.085$.

\begin{figure}[!ht]
	\centering
	\includegraphics[width=0.4\textwidth]{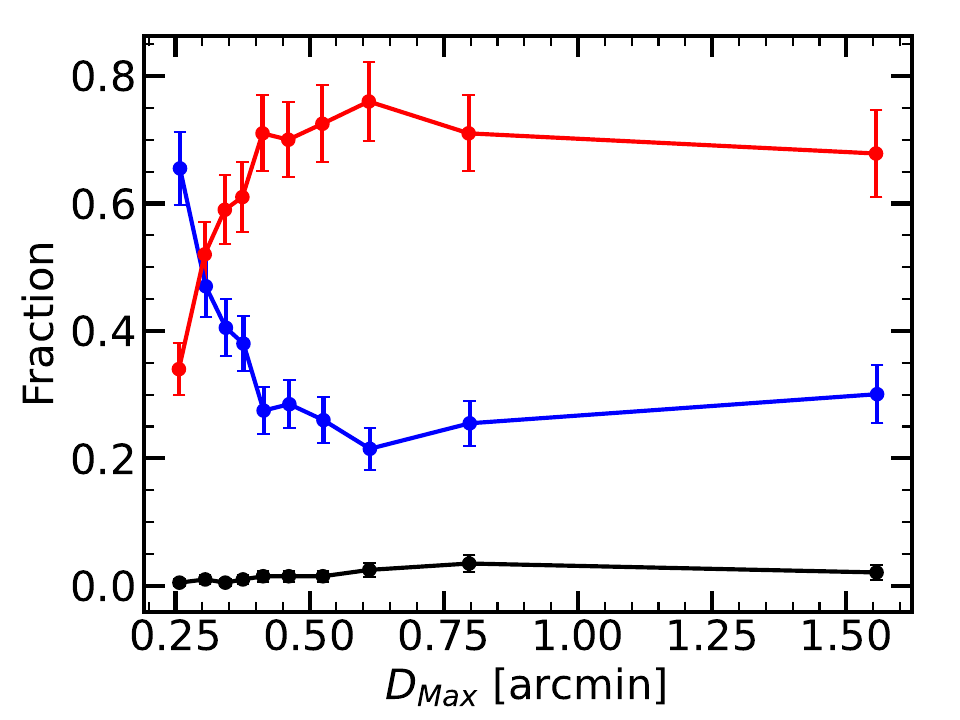}
	\includegraphics[width=0.4\textwidth]{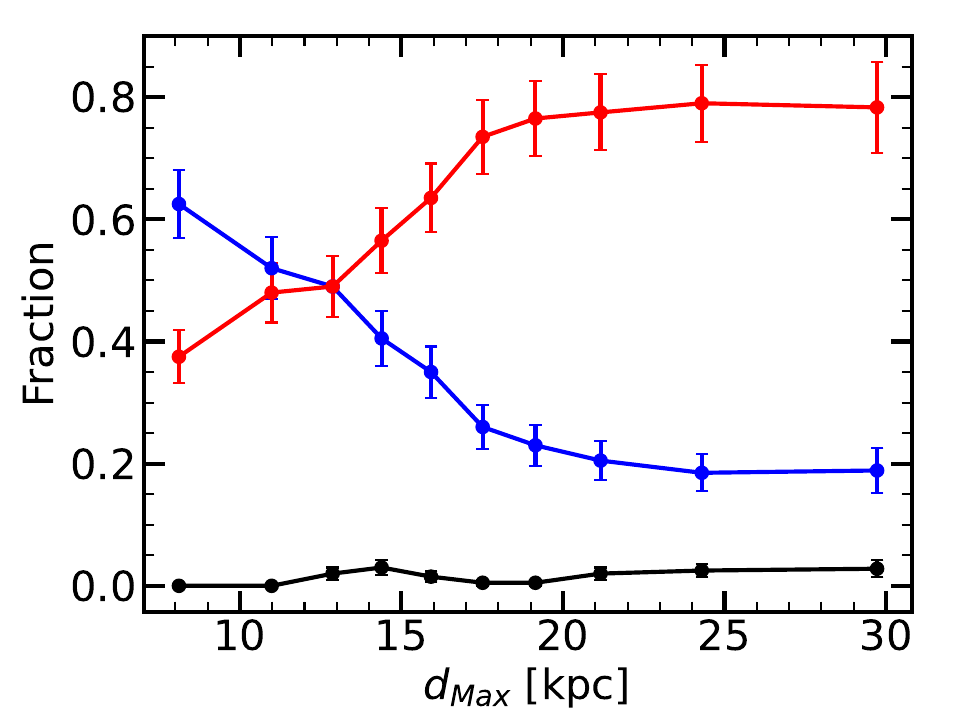}
	\caption{The galaxy morphology statistics under different scales. Left: the result corresponding to the angular scale. Right: the result corresponding to physical scale. Shown are flocculent galaxies (blue), multiple-arm galaxies (red), and grand-design galaxies (black).}
	\label{fig:size_morpholoty}
\end{figure}

The statistical results of spiral arm morphologies as a function of galaxy scale are presented in Figure~\ref{fig:size_morpholoty}, using a similar binning method as described above. At smaller galaxy sizes ($D_{\mathrm{max}} < 0.45'$), there is a clear trend of increasing proportions of flocculent galaxies and decreasing proportions of multiple-arm galaxies. This trend becomes more pronounced as galaxy size further decreases. 
When the galaxy size exceeds $0.45'$, the morphological distribution stabilizes, with multiple-arm galaxies comprising a dominant fraction (approximately 70\%). Converting angular size to physical scale using redshift further reinforces this trend, showing that when a major axis of galaxy exceeds 18~kpc, the proportion of multiple-arm galaxies can reach up to 80\%. 

We have statistically analyzed the distribution of spiral arm morphologies among MC subsample across different Hubble types (see Table~\ref{tab:hubble_morphology}). However, the number of galaxies available for certain Hubble subtypes remains limited, making it difficult to draw robust conclusions or perform more detailed analyses for these specific categories. We anticipate that future studies with larger samples will allow for a more comprehensive investigation. In this paper, we simply present the current results as an initial exploration.

\begin{table*}[!ht]
	\small
	\centering
	\caption{Morphological statistics of spiral arm morphology for different Hubble type in the morphologically consistent sample}
	\label{tab:hubble_morphology}
	\begin{tabular}{|l|cccccc|} 
		\hline
		  & SABb & SABbc & SABc & SBb & SBbc & SBc \\
		\hline
		F &  5 (62\%) &  5 (62\%) &  5 (28\%) & 284 (38\%) &  15 (23\%) &  69 (34\%) \\
		M &  3 (38\%) &  3 (38\%) & 13 (72\%) & 444 (60\%) &  47 (73\%) & 135 (66\%) \\
		G &  0 ( 0\%) &  0 ( 0\%) &  0 ( 0\%) &  13 ( 2\%) &   2 ( 3\%) &   0 ( 0\%) \\
		\hline
	\end{tabular}
\end{table*}

\subsection{Number of inner arms}\label{sec3_3}

As early as the 20th century, galaxy atlases such as \textit{The Atlas of Galaxies}~\citep{Sandage_Bedke1988}, \textit{The Revised Shapley-Ames Catalog of Bright Galaxies}~\citep{Sandage_Tammann1981}, and \textit{The Hubble Atlas}~\citep{Sandage1961} already indicated that the majority of spiral galaxies possess two inner arms. This observation laid the foundation for the widely held notion in subsequent morphological classifications that both grand-design and multiple-arm galaxies generally exhibit a symmetric two-arm structure in their inner regions~\citep{Elmegreen_Elmegreen1982,Elmegreen_Elmegreen1987}. A detailed statistical analysis of 3,958 galaxies by \citet{Wei+2024} further confirmed this trend, showing that approximately 82\% of spiral galaxies possess two inner arms.

In this study, we focus on MW-like barred spiral galaxies. The MW itself is considered to possess distinct multiple spiral arms~\citep{Xu+2013,Reid+2019,Xu+2021}, resembling typical multiple-arm galaxies~\citep{Xu+2023}. Accordingly, we statistically analyze the number of inner arms in MW-like multiple-arm galaxies, with the results summarized in Table~\ref{tab:innerarms}.

Following the classification scheme of \citet{Elmegreen_Elmegreen1982,Elmegreen_Elmegreen1987}, galaxies with only one inner arm are classified as flocculent (class 4). We adopt this system, and thus exclude galaxies with a single inner arm from our statistics in Table~\ref{tab:innerarms}. In our sample, only one galaxy (NVSS J134027+044627, based on the SDSS $g$-band image) was classified by one author as having five inner arms. For simplicity, we categorize the number of inner arms as 2, 3, or 4+ ($\geqslant 4$).

As shown in Table~\ref{tab:innerarms}, the majority of multiple-arm galaxies exhibit two inner arms ($\sim$90 $\pm$ 3\%), followed by three arms ($\sim$9 $\pm$ 2\%), with a very small fraction ($<$1\%) displaying four or more. These results are consistent with \citet{Wei+2024}, who found that 87\% of barred multiple-arm galaxies possess two inner arms when analyzed separately. It is important to note that this study focuses specifically on the number of inner arms, which may differ from previous works that classify spiral structure based on the total number of arms observed in the entire disk~\citep[e.g.,][]{Hart+2016,Savchenko+2020}, resulting in different proportions for two-arm, three-arm, and galaxies with four or more spiral arms.

\begin{table*}[!ht]
	\small
	\centering
	\caption{Distribution of the numbers of inner arms in MW-like barred spiral galaxies}
	\label{tab:innerarms}
	\begin{tabular}{|l|cc|cc|cc|cc|cc|cc|} 
		\hline
		  &   \multicolumn{6}{c|}{author (A)} &    \multicolumn{6}{c|}{author (B)} \\
		  \hline
		  &   \multicolumn{2}{c|}{2}  &  \multicolumn{2}{c|}{3}   & \multicolumn{2}{c|}{4+}  &   \multicolumn{2}{c|}{2}  &  \multicolumn{2}{c|}{3}   & \multicolumn{2}{c|}{4+}  \\
		  & Number & \% & Number & \% & Number & \% & Number & \% & Number & \% & Number & \% \\
		\hline
		DSS blue & 170 & 88 & 24 & 12 & 0 & 0    & 201 & 87 & 28 & 12 & 1 & $<$1 \\
		DSS red  & 155 & 92 & 13 &  8 & 1 & $<$1 & 172 & 90 & 18 &  9 & 2 & $<$1 \\
		SDSS $g$   & 756 & 93 & 46 &  6 & 8 & $<$1 & 793 & 93 & 55 &  6 & 4 & $<$1 \\
		SDSS $r$   & 734 & 91 & 65 &  8 & 4 & $<$1 & 790 & 92 & 62 &  7 & 3 & $<$1 \\
		SDSS $i$   & 697 & 93 & 52 &  7 & 3 & $<$1 & 742 & 93 & 57 &  7 & 2 & $<$1 \\
		\hline
	\end{tabular}
\end{table*}

Table~\ref{tab:innerarms_vs_band} presents the variations in the number of inner arms across different bands. In this discussion, we focus on multiple-arm galaxies in MC subsample (a total of 645 galaxies). Since the identification of inner-arm numbers involves a degree of subjectivity, two authors independently performed the classifications. As minor discrepancies may arise across different bands, we retain both sets of classifications in the subsequent statistics, effectively doubling the sample size to account for this uncertainty.
From Table~\ref{tab:innerarms_vs_band}, several trends can be identified. Galaxies with two inner arms exhibit extremely high stability across all band transitions (DSS blue $\rightarrow$ DSS red, SDSS $g$ $\rightarrow$ SDSS $r$, SDSS $r$ $\rightarrow$ SDSS $i$, SDSS $g$ $\rightarrow$ SDSS $i$), with the vast majority preserving their two-arm structure. Galaxies with three inner arms show more dynamic behavior, with transitions either toward a two-arm structure or retaining the original morphology. In contrast, multiple-arm galaxies with four or more inner arms are exceedingly scarce in all bands and display substantial morphological variability across different wavelengths.

\begin{table*}[!ht]
	\small
	\centering
	\caption{Differences in the number of inner arms across bands}
	\label{tab:innerarms_vs_band}
	\begin{tabular}{|l|c|c|c|c|c|c|c|c|c|} 
		\hline
		   & $2 \rightarrow 2$ & $2 \rightarrow 3$ & $2 \rightarrow 4+$ & $3 \rightarrow 2$ & $3 \rightarrow 3$ & $3 \rightarrow 4+$ & $4+ \rightarrow 2$ & $4+ \rightarrow 3$ & $4+ \rightarrow 4+$  \\
		\hline
		DSS blue $\rightarrow$ DSS red  & 231 &  5 & 1 & 16 & 15 & 0 & 0 & 0 & 0 \\
		\hline
		SDSS $g$ $\rightarrow$ SDSS $r$     & 872 & 59 & 1 & 34 & 42 & 2 & 3 & 6 & 3 \\
		SDSS $r$ $\rightarrow$ SDSS $i$     & 872 & 36 & 1 & 57 & 50 & 0 & 4 & 1 & 1 \\
		SDSS $g$ $\rightarrow$ SDSS $i$     & 884 & 48 & 0 & 45 & 32 & 1 & 4 & 7 & 1 \\
		\hline
	\end{tabular}
\end{table*}

\begin{figure}[!ht]
	\centering
	\includegraphics[width=0.65\textwidth]{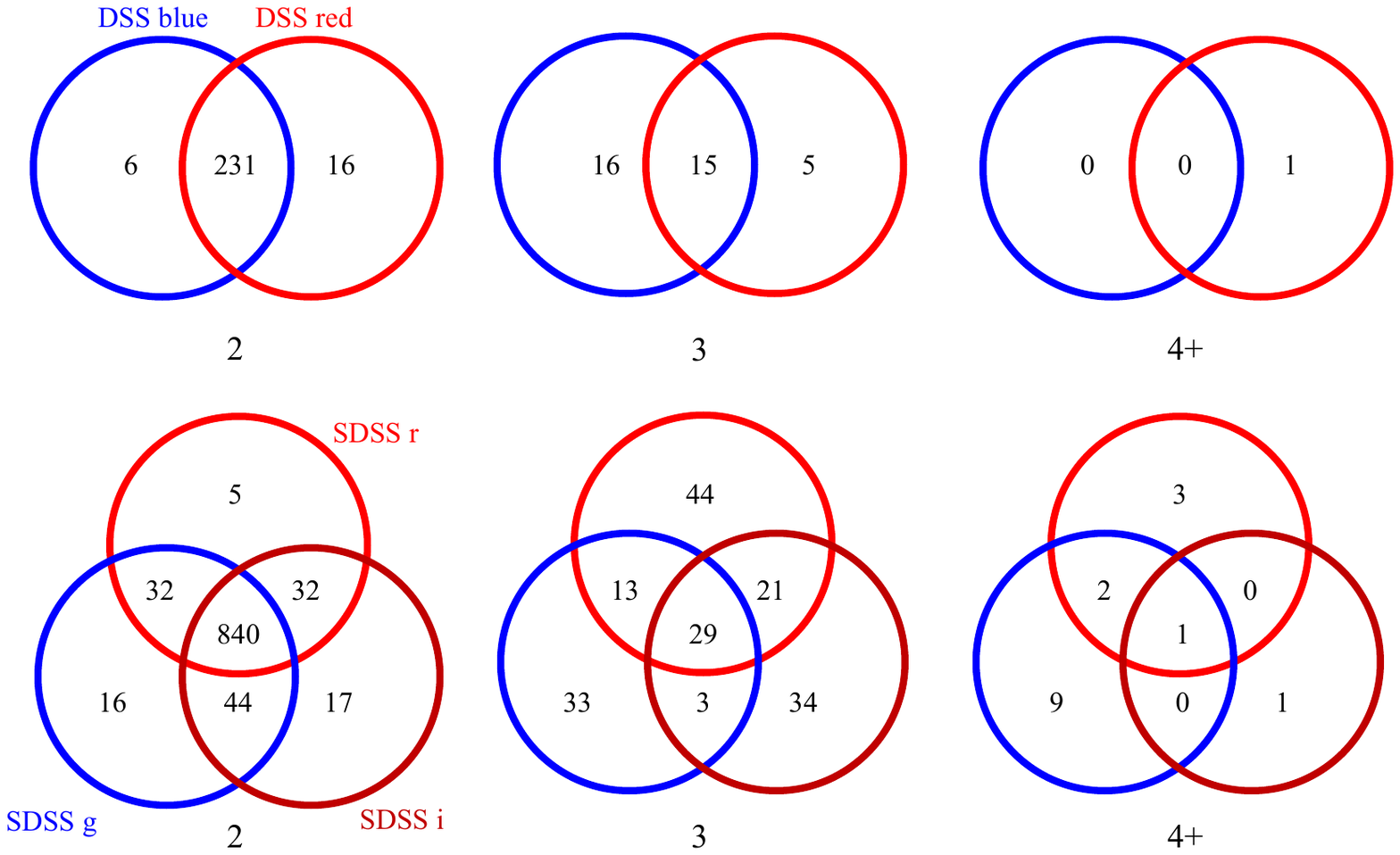}
	\caption{Distribution of the numbers of inner arms in multiple-arm galaxies across different bands. Top: DSS blue and DSS red bands. Bottom: SDSS $g$-, $r$-, and $i$-bands. From left to right, the images systematically display: galaxies with two inner arms, three inner arms, and four or more inner arms. The intersecting parts indicate that the galaxies exhibit the equal number of inner arms in these bands.}
	\label{fig:innerarm_vs_band}
\end{figure}

Figure~\ref{fig:innerarm_vs_band} illustrates the distribution of inner arm counts for galaxies observed across different bands. Taking the bottom-left panel as an example: if a galaxy is classified as having two inner arms in the SDSS $g$-band, then in 90\% of cases (840/932) it is also classified as having two inner arms in the other bands, and in only 2\% of cases (16/932) do the other bands yield a different classification. In contrast, galaxies classified as having three or more inner arms in a given band are far less likely to retain the same count across other bands, suggesting that such classifications are often band-specific.

Within the sample of multiple-arm galaxies, 83\% exhibit two inner arms consistently across all bands, 3\% consistently exhibit three inner arms, and only a single case retains four inner arms in all bands. In summary, multiple-arm galaxies with two inner arms display a high degree of cross-band stability in their inner arm count, whereas classifications with $\geqslant 3$ inner arms show markedly lower consistency.

\subsection{Bifurcation}\label{sec3_4}

For multiple-arm galaxies, the spiral arms are generally less clear, long, and continuous than in grand-design systems. Local kinks, discontinuities, bifurcations, and contamination from background objects often cause the arms to split or appear as isolated segments, thereby increasing the difficulty of automated spiral arm fitting~\citep[e.g.,][]{Davis_Hayes2014}. Such irregularities typically occur in the outer regions, whereas the inner regions of multiple-arm galaxies usually host only two or a few spiral arms~\citep{Elmegreen_Elmegreen1982,Elmegreen_Elmegreen1987}.

Understanding the transition from relatively regular inner arms to multiple outer arms is thus essential for characterizing the structure of these galaxies. Early work by \citet{Elmegreen_Elmegreen1995} based on 173 spiral galaxy images showed that bifurcations or broadenings of the inner arms typically occur at $\sim$0.5\,$R_{25}$ or within twice the bar radius, where $R_{25}$ is the radius at a surface brightness of 25\,mag\,arcsec$^{-2}$. More recently, \citet{Wei+2024} statistically analyzed over 1,800 spiral galaxies with distinct bifurcation points and found that 91$\pm$3\% of these points lie within 0.2--0.6\,$R$, where $R$ denotes the galaxy edge in the image.

\begin{figure}[!ht]
	\centering
	\includegraphics[width=0.9\textwidth]{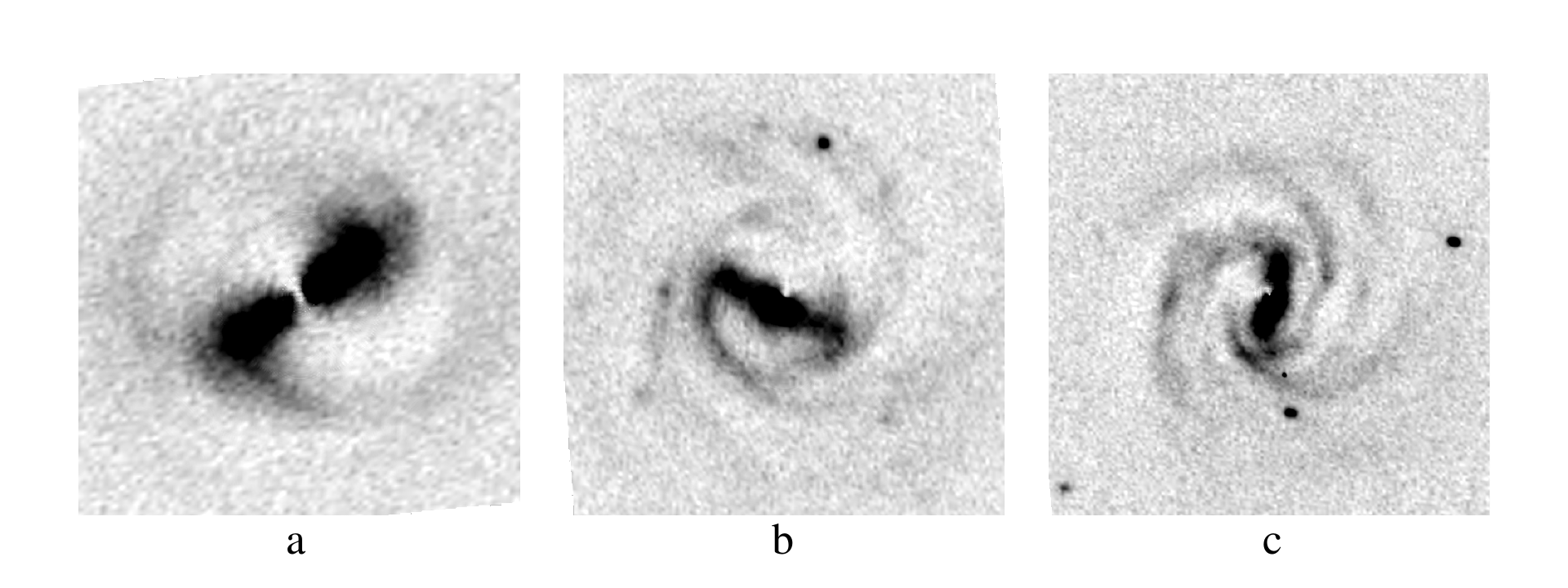}
	\caption{Typical multiple-arm galaxies. These images show the luminance compared to background. Panel a: LEDA 2404450 in SDSS $r$-band, panel b: UGC 4369 in SDSS $r$-band, panel c: UGC 855 in SDSS $g$-band.}
	\label{fig:multiarm_typical}
\end{figure}

Figure~\ref{fig:multiarm_typical} shows representative examples of multiple-arm galaxies in our sample.  
A small fraction of systems display identical numbers of inner and outer arms yet lack long, continuous spiral structures, with arms instead appearing as short or disconnected segments (Figure~\ref{fig:multiarm_typical}a).  
Across all bands, such cases are found 94 times in the 2,477 samples (102 according to the other author), corresponding to $\sim$4\% of the total; when restricting to consensus classifications, this number decreases to 36.

In contrast, the majority of multiple-arm galaxies display a discrepancy between inner and outer arm counts, most commonly two inner arms transitioning to multiple outer arms.  
Two primary transition patterns are identified:  
(1) outer arms lacking clear continuity with the inner structure, i.e., not branching from inner arms (Figure~\ref{fig:multiarm_typical}b);  
(2) outer arms directly connected to the inner structure via bifurcations (Figure~\ref{fig:multiarm_typical}c).  

Statistically, 1,719 (1,332) galaxies match type~(1), and 664 (1,043) match type~(2), where numbers outside and inside the parentheses represent the independent classifications of the two authors.
Differences between authors likely arise from galaxies exhibiting both bifurcations and isolated outer segments.  
Consensus classifications are reached for 1,086 galaxies without branching and 475 with distinct bifurcations, while 916 remain disputed.

We note that our statistics consider only the first change in arm number from the galactic center outward; bifurcations occurring farther in the outer disk are excluded.  
Among the 475 bifurcation cases, the distribution across bands is: 52 (DSS blue), 30 (DSS red), 169 (SDSS $g$), 138 (SDSS $r$), and 86 (SDSS $i$).  
If disputed cases are also included, the fraction of galaxies in which bifurcation points can be identified becomes $42\pm7$\% (DSS blue), $35\pm9$\% (DSS red), $38\pm7$\% (SDSS $g$), $34\pm7$\% (SDSS $r$), and $29\pm8$\% (SDSS $i$).  
While the overall spiral morphology and inner arm count show no significant cross-band variation, the occurrence of bifurcation points exhibits a clear dependence on observational band, with a higher prevalence in bluer wavelengths.

\begin{figure}[!ht]
	\centering
	\includegraphics[width=0.4\textwidth]{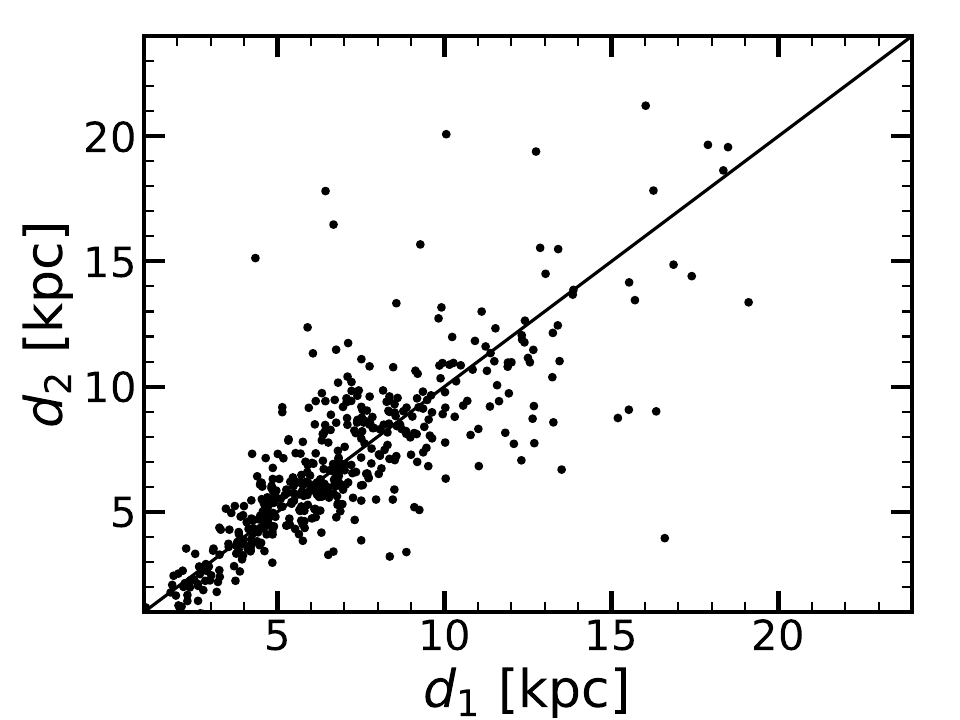}
	\includegraphics[width=0.4\textwidth]{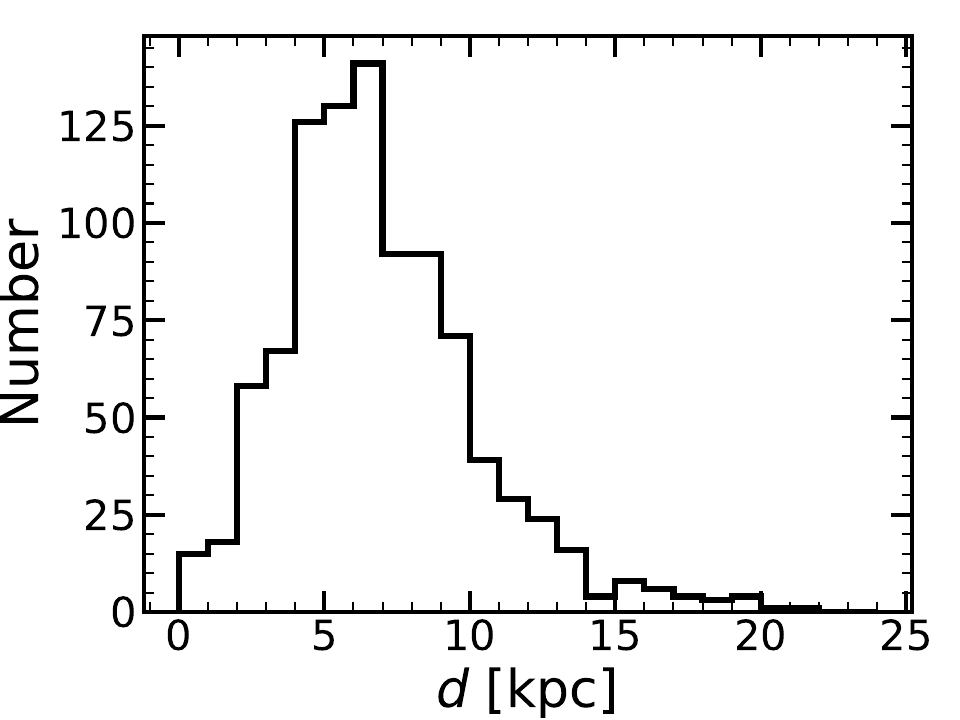}
	\caption{Position distribution of bifurcation points closest to the galactic center}
	\label{fig:bifurcation}
\end{figure}

For the 475 galaxies in our sample with distinct inner-to-outer bifurcations, the positions of the bifurcation points closest to the galactic center are shown in Figure~\ref{fig:bifurcation}. The left panel shows the measurements from the two authors, which exhibit a high level of consistency. Specifically, 68\% of the differences between their measurements are smaller than 1.35 kpc. The right panel combines both authors' results and shows that these bifurcation points are predominantly located at galactocentric distances of $\sim$4--7\,kpc.

\section{Comparison of the statistical results and the MW}
\label{sec4}

The Milky Way has been considered a spiral galaxy since at least the 1850s \citep{Alexander1852}, but its spiral structure is notoriously difficult to observe due to our edge-on vantage point and heavy dust extinction. In the 1950s, high-mass stars (OB stars) and H{\sc II} regions were first used to trace local spiral arm segments \citep{Morgan+1952,Morgan+1953}. Building on these observations, \cite{Georgelin_Georgelin1976} combined optical data of young stars with radio observations of H{\sc I} clouds and H{\sc II} regions to propose the classical four-arm model of the MW. Despite large uncertainties in kinematic distances, this model remains the foundation of the ``standard'' view of our Galactic spiral structure, subsequently refined by later works \citep[e.g.,][]{Taylor_Cordes1993,Levine+2006}.

Recent maser parallax measurements have enabled precise localization of high-mass star-forming regions. Using $\sim$200 high-mass star-forming region masers, \cite{Reid+2019} produced the most detailed map of the spiral arms of MW to date, revealing four main arms along with additional segments and spurs. While the MW has long been classified as a grand-design galaxy, the presence of the Local Arm \citep{Xu+2013,Xu+2016} and evidence of localized arm fragmentation, kinks, and inhomogeneities \citep{Reid+2019,Xu+2021} suggest it departs significantly from the global symmetry of typical grand-design spirals, and may instead belong to the multiple-arm category.

Based on the arm segment of \cite{Reid+2019} and classical Cepheid observations, \cite{Minniti+2021} proposed that the Sagittarius-Carina Arm and Perseus Arm intersect, and that the Perseus Arm connects inward with the Norma Arm, producing an inner two-arm structure formed by the Norma and Scutum Arms. In contrast, \cite{Xu+2023} synthesized maser parallaxes, 3D positions of massive stars from {\it Gaia} DR3, and spiral arm tangent points to propose that the MW is a multiple-arm galaxy with two inner arms transitioning to multiple outer arms. Their model shows that the traditional Norma and Scutum Arms are actually a single continuous structure; the Perseus Arm, extending toward the Galactic center, forms a symmetric inner two-arm pattern with the Norma Arm; and, for the first time, symmetrical bifurcation points were identified at 5.0--6.5~kpc from the Galactic center, beyond which multiple outer arms emerge.

Through the statistical analysis presented in this study, we find that the MW spiral arm structure proposed by \cite{Xu+2023} aligns well with the morphology of typical multiple-arm galaxies, for the following reasons.

\textit{(1) Consistency between young and old stellar tracers.} The \textit{Spitzer} GLIMPSE survey \citep{Benjamin+2005,Churchwell+2009} revealed that the MW possesses only two main arms in its inner region when traced in the near-infrared, corresponding to the old stellar population. The new spiral arm structure traced by young objects in \cite{Xu+2023} likewise exhibits two prominent inner arms. Moreover, \cite{Nakanishi+2024} found that the Galactic epicyclic frequency profile supports a two-arm pattern in the inner Galaxy and a four-arm pattern in the outer regions. Our results further show that spiral galaxy morphology is generally consistent across different wavelength bands. Likewise, extragalactic studies \citep{Buta+2015,Elmegreen+2011} have demonstrated that the morphology traced by older stellar populations in the near-infrared spatially coincides with that traced by younger stellar populations in the optical. Therefore, the inner two-arm structure in the \cite{Xu+2023} model, which unifies young and old stellar tracers, appears to represent a more universal configuration.

\textit{(2) Existence of spiral arm bifurcation points.} \cite{Xu+2023} reported that the two inner arms bifurcate into four outer arms at a Galactocentric radius of 5.0--6.5~kpc, consistent with the innermost bifurcation points observed in the majority of galaxies in our sample. Additional studies also support such bifurcation features: recent maser observations suggest possible intersections between the Sagittarius--Carina and Perseus Arms \citep{Bian+2024}, while two adjacent tangent points in the fourth quadrant indicate potential bifurcation scenarios \citep{Hou_Han2015,Minniti+2021}. Furthermore, our statistical results show that bifurcation points are more frequently detected in the inner arms of multiple-arm galaxies in bluer (younger) bands, implying that masers in high-mass star-forming regions and other young tracers in the MW may more effectively delineate these bifurcations.

Combining the above points, the MW model proposed by \cite{Xu+2023} exhibits strong consistency with the statistical properties of spiral galaxies. If correct, this model would revise our view of the MW from an atypical four-arm system \citep{Georgelin_Georgelin1976,Taylor_Cordes1993} to a typical multiple-arm galaxy with two inner arms transitioning to multiple outer arms.

\section{Conclusions}
\label{sec5}

This study systematically analyzes the morphological characteristics of 1,738 nearby MW-like barred spiral galaxies (inclination $<60^\circ$, redshift $Z < 0.048$) using multi-band optical data from SDSS and DSS. Following detailed investigation of spiral arm morphology, inner structures, and bifurcation properties, we present the following conclusions:

\begin{itemize}
	\item[1] MW-like galaxies predominantly exhibit multiple-arm structures ($\sim 50$--$60$\%), whereas grand-design galaxies constitute only about 4\%. For large-scale galaxies ($d_{\rm max} \geqslant 18~\mathrm{kpc}$), the fraction of multiple-arm galaxies reaches as high as 75\%.
	\item[2] The morphological consistency of spiral arm structures across different bands remains as high as 80\%.
	\item[3] In multiple-arm galaxies, the inner arms are dominated by two-arm structures ($90 \pm 3\%$), followed by three-arm structures ($9 \pm 2\%$), while four or more arms are extremely rare ($<1\%$). The inner-arm number remains consistent across different observational bands at the $\sim 80\%$ level.
	\item[4] Certain multiple-arm galaxies exhibit bifurcation points where the inner two arms transition into outer multiple arms. These bifurcation points are predominantly clustered within 4--7~kpc from galactic centers, and occur more frequently in bluer bands: $42 \pm 7$\% (DSS blue), $38 \pm 7$\% (SDSS $g$), $35 \pm 9$\% (DSS red), $34 \pm 7$\% (SDSS $r$), and $29 \pm 8$\% (SDSS $i$). Considering that the DSS blue ($\lambda_{\rm eff} \approx 4800$~\AA) and SDSS $g$ ($\lambda_{\rm eff} \approx 4770$~\AA) bands trace younger stellar populations compared to the DSS red ($\lambda_{\rm eff} \approx 6500$~\AA), SDSS $r$ ($\lambda_{\rm eff} \approx 6230$~\AA), and SDSS $i$ ($\lambda_{\rm eff} \approx 7630$~\AA), these results indicate that bifurcation points are more distinct in bluer (younger) wavelengths, supporting the interpretation that such features preferentially trace regions of ongoing or recent star formation.
\end{itemize}

Taken together, the MW model proposed by \cite{Xu+2023}---featuring an inner two-arm structure, bifurcation points within 5--6.5~kpc, and outer multiple-arm extensions---is in remarkable agreement with the statistical properties of MW-like galaxies presented here. This suggests that, contrary to the long-standing four-arm morphology, the multiple-arm MW is morphologically consistent with a typical spiral galaxy.

\begin{acknowledgements}
We thank the anonymous referee for the useful suggestions and comments.
This work was funded by the National SKA Program of China (Grant No. 2022SKA0120103) and the Key Laboratory for Radio Astronomy. Z.H.L. thanks the support of the NSFC grant No. 12403077. Y.J.L. thanks the support of the NSFC grant No. 12203104. C.J.H. acknowledges support from the National Postdoctoral Program for Innovative Talents of the Office of China Postdoc Council (grant No. BX20240414) and the NSFC grant No. 12403041. 
This research has made use of the SIMBAD database and the VizieR catalog, operated at CDS,Strasbourg, France.
This work has made use of data from SDSS-V.
Funding for the Sloan Digital Sky Survey V has been provided by the Alfred P. Sloan Foundation, the Heising-Simons Foundation, the National Science Foundation, and the Participating Institutions. 
SDSS acknowledges support and resources from the Center for High-Performance Computing at the University of Utah. 
SDSS telescopes are located at Apache Point Observatory, funded by the Astrophysical Research Consortium and operated by New Mexico State University, and at Las Campanas Observatory, operated by the Carnegie Institution for Science. 
The SDSS web site is \url{www.sdss.org}. This research has made use of the Digitized Sky Survey, produced at the Space Telescope Science Institute under U.S. Government grant NAG W-2166, based on photographic data from the Palomar and UK Schmidt telescopes.

\end{acknowledgements}

\appendix

\section{Statistical results of MW-like Galaxies}\label{seca}

This chapter presents statistical result of galaxies. Column 1 the lists galaxy name, column 2 shows the Hubble type, column 3 shows redshift, and column 4 and 5 list the major and minor axis of galaxies in arcminutes, respectively. The information from column 2 to 5 is all sourced from SIMBAD. Column 6 indicates the band used for statistics. Column 7 and 8 list the spiral arm morphologies of galaxies as identified by different authors, where F, M, and G denote flocculent, multiple-arm, and grand-design galaxies, respectively. Column 9 and 10 show the number of inner arms of galaxies, as statistically obtained by different authors, specifically for multiple-arm and grand-design galaxies. Column 11 and 12 indicate whether different authors agree that the number of outer arms matches the number of inner arms (Yes/No). Column 13 and 14 denote whether different authors observe a clear bifurcation point between the inner and outer arms of galaxies (Yes/No). Column 15 and 16 provide the distance from the bifurcation point to the galactic center, measured in kpc, if such a bifurcation point exists. For the full table, please refer to the online version. 

\begin{table*}[!ht]
  \renewcommand{\thetable}{A1}
  \tiny
  \caption{Statistical results of MW-like Galaxies}
  \label{tab:galaxies}
  \hspace{-1.3cm}
  \begin{tabular}{|l|c|c|c|c|c|c|c|c|c|c|c|c|c|c|c|} 
    \hline
      Name & Hubble type & redshift & $D_{Max}$ & $D_{Min}$ & band & Type (A) & Type (B) & $N_{i,A}$ & $N_{i,B}$ & Same (A) & Same (B) & B (A) & B (B) & $B_{i,A}$ & $B_{i,B}$  \\
      (1) & (2) & (3) & (4) & (5) & (6) & (7) & (8) & (9) & (10) & (11) & (12) & (13) & (14) & (15) & (16) \\
    \hline
    NGC  5832            & SBb   & 0.001 &  3.70 &  2.20 & DSS blue & F & F & 0 & 0 & N & N & N & N &  0.0 & 0.0 \\
    NGC  5832            & SBb   & 0.001 &  3.70 &  2.20 & DSS red  & F & F & 0 & 0 & N & N & N & N &  0.0 & 0.0 \\
    M  83                & SABc  & 0.002 & 13.80 & 12.88 & DSS blue & F & M & 0 & 3 & N & N & N & N &  0.0 & 0.0 \\
    M  83                & SABc  & 0.002 & 13.80 & 12.88 & DSS red  & F & M & 0 & 2 & N & N & N & N &  0.0 & 0.0 \\
    NGC  3344            & SABbc & 0.002 &  6.76 &  6.46 & SDSS g   & M & M & 2 & 2 & N & N & Y & Y &  2.1 & 1.2 \\
    NGC  3344            & SABbc & 0.002 &  6.76 &  6.46 & SDSS r   & M & M & 2 & 2 & N & N & Y & Y &  2.1 & 1.2 \\
    NGC  3344            & SABbc & 0.002 &  6.76 &  6.46 & SDSS i   & M & M & 2 & 2 & N & N & Y & Y &  2.0 & 1.3 \\
    \hline
  \end{tabular}
\end{table*}

\section{Consistency Between two authors}\label{secb}
\subsection{Consistency in Morphology}

Table~\ref{tab:diference_vs_person} presents the differences in spiral arm morphology classifications between the two authors across different bands. The overall agreement rates are 86\%, 89\%, 85\%, 84\%, and 85\% for the DSS blue, DSS red, and SDSS $g$-, $r$-, and $i$-bands, respectively. Most disagreements occur between adjacent morphological classes, such as between flocculent and multiple-arm, or between multiple-arm and grand-design galaxies. 

These classification differences are comparable to those reported in a previous study~\citep{Wei+2024}, where the same set of galaxies was reclassified after a one-year interval. As the sample size increases, the impact of such discrepancies on the overall statistical results becomes negligible. For samples containing thousands of galaxies, classification inconsistencies between independent inspections alter the overall morphological fractions by only $\sim$2\%.

\begin{table*}[!ht]
    \renewcommand{\thetable}{B1}
    \tiny
	\caption{Differences in spiral arm morphology statistics between different authors}
	\label{tab:diference_vs_person}
	\begin{tabular}{|c|ccc|ccc|ccc|ccc|ccc|} 
		\hline
		   & \multicolumn{3}{c|}{DSS blue} & \multicolumn{3}{c|}{DSS red} & \multicolumn{3}{c|}{SDSS $g$} & \multicolumn{3}{c|}{SDSS $r$} & \multicolumn{3}{c|}{SDSS $i$}  \\
		\hline
		   &  F (A) &  M (A) &  G (A) &  F (A) &  M (A) &  G (A) & F (A) & M (A) & G (A) & F (A) & M (A) & G(A)  & F (A) & M (A) & G (A) \\
		\hline
		F (B) & 571 & 180 &   4 & 659 & 175 &   3 & 389 &  63 &   0 & 380 &  60 &   2 & 445 &  59 &   1 \\ 
		M (B) & 181 & 724 &  21 & 148 & 681 &  26 &  82 & 731 &  39 &  86 & 726 &  43 &  73 & 676 &  52 \\ 
		G (B) &   1 &  26 &  30 &   1 &  19 &  26 &   0 &  16 &  22 &   2 &  17 &  26 &   0 &  17 &  19 \\ 
		\hline
	\end{tabular}
\end{table*}

\subsection{Consistency in inner arm classifications}

The consistency of inner-arm classifications between the two authors is high across all five bands, with agreement rates of 97\%, 96\%, 99\%, 98\%, and 99\% in the DSS blue, DSS red, SDSS $g$-, $r$-, and $i$-bands, respectively (Table~\ref{tab:diference_vs_innerarms}).
Cases with four or more inner arms are rare in all bands, and their classification agreement is noticeably lower, likely because the crowded and irregular arrangement of numerous inner arms can lead to misidentification as two- or three-arm systems.

\begin{table*}[!ht]
    \renewcommand{\thetable}{B2}
	\small
	\caption{Differences in the number of inner arms counted by the authors}
	\label{tab:diference_vs_innerarms}
	\begin{tabular}{|l|ccc|ccc|ccc|ccc|ccc|} 
		\hline
		   &   2  &  3   & 4+ &   2  &  3   & 4+ &   2  &  3   & 4+ &   2  &  3   & 4+ &   2  &  3   & 4+  \\
		\hline	
		   & \multicolumn{3}{c|}{DSS blue} & \multicolumn{3}{c|}{DSS red}& \multicolumn{3}{c|}{SDSS $g$}& \multicolumn{3}{c|}{SDSS $r$}& \multicolumn{3}{c|}{SDSS $i$}\\
		\hline	
		2  & 159 &  2 & 0 & 141 &  3 & 1 & 672 &  5 & 0 & 655 &  3 & 0 & 617 &  4 & 1 \\
		3  &   3 & 21 & 0 &   1 & 12 & 0 &   1 & 45 & 0 &   6 & 56 & 2 &   3 & 48 & 0 \\
		4+ &   0 &  0 & 0 &   1 &  0 & 0 &   3 &  1 & 4 &   3 &  0 & 1 &   2 &  0 & 1 \\
		\hline
	\end{tabular}
\end{table*}

\bibliographystyle{raa}
\bibliography{galaxies}

\label{lastpage}
\end{document}